\documentclass[12pt]{amsart}

\usepackage{amsmath,amssymb,amsthm,a4,bm,epsfig,color,graphics}

\newtheorem{theorem}{Theorem}

\newtheorem*{remark*}{Remark}
\newtheorem*{remarks*}{Remarks}

\newcommand{\bbR}{{\mathbb R}}

\newcommand{\cB}{{\mathcal B}}
\newcommand{\cH}{{\mathcal H}}

\newcommand{\ap}{a^{(p)}}
\newcommand{\XN}{X_N^{(p)}}
\newcommand{\X}{X^{(p)}}
\newcommand{\YN}{Y_N^{(p)}}
\newcommand{\YNl}{Y_N^{(p), <\beta}}
\newcommand{\YNg}{Y_N^{(p), \ge \beta}}

\begin{document}

\title[]{A remark on the mean-field dynamics of many-body bosonic systems with random interactions}

\author[]{Walid K. Abou Salem}

\address{Department of Mathematics, University of Toronto, Toronto, Ontario, Canada M5S 2E4 \\ E-mail: walid@math.utoronto.ca} 

\maketitle

\begin{abstract}
The mean-field limit for the dynamics of bosons with random interactions is rigorously studied. It is shown that, for interactions that are almost surely bounded, the many-body quantum evolution can be replaced in the mean-field limit by a single particle nonlinear evolution that is described by the Hartree equation. This is an Egorov-type theorem for many-body quantum systems with random interactions.
\end{abstract}

%%%%%%%%%%%%%%%%%%%%%%%%%%%%%%%%%%%%%%%%%%%%%%%%%%%%%%
%INTRODUCTION
%%%%%%%%%%%%%%%%%%%%%%%%%%%%%%%%%%%%%%%%%%%%%%%%%%%%%%

\section{Introduction}

This work is a modest contribution to the mathematical theory of the mean-field limit for bosons with random interactions.  There has been substantial developments in the study of the mean-field dynamics of bosons with deterministic interactions. Early results were proven by Hepp in \cite{He}, see also \cite{GiVe}. A different approach based on the reduced density matrix was developed  in \cite{Sp} and was substantially 
extended to more general potentials and to the derivation of the Gross-Pitaevskii equation in \cite{EY}, \cite{BGM}, \cite{BEGMY}, \cite{ESY1}, \cite{ESY2}. Recently, a new approach was developed in \cite{FGS}, which gives convergence estimates in the mean-field limit that are uniform in Planck's constant $\hbar,$ see also \cite{FKP}.

While the mean-field dynamics for bosons with deterministic interactions has attracted considerable interest, the question of the mean-field dynamics of bosons with random interactions has not been addressed, yet. Many-body bosonic systems with random interactions are relevant to concrete physical systems, such as inhomogeneous nonlinear optical media, or Bose-Einstein experiments where irregular fluctuations in currents inside conductors close to the condensate induce via Feshbach resonances inhomogeneous interactions between the bosons, see \cite{Wildermuth} for a description of the latter; also \cite{AG} and references therein. Here, we give a simple recepy for extending the deterministic mean-field analysis to the case of random interactions (and in the presence of a random potential).

\subsection{The model}\label{sec:Model}

Consider the probability triple $(\Omega, {\mathcal F}, {\mathbf P}),$ such that the probability space $\Omega$ has a generic point $\omega$ and is endowed with measure $\mu.$ Define on this space the random field 
\begin{equation*}
v(x,\omega) \ \ : \ \ {\mathbb R}^3\times\Omega \rightarrow {\mathbb R}, 
\end{equation*} 
such that $v$ is measurable in $x\in {\mathbb R}^3$ and $\omega\in\Omega,$ and is almost surely in $L^\infty({\mathbb R}^3),$ i.e. there exists $\Omega_0\subset \Omega$ such that $\mu(\Omega_0)=1$ and, for all $\omega\in\Omega_0,$ $v(\cdot, \omega)\in L^\infty({\mathbb R}^3).$ A concrete example of $v$ that satisfies the above conditions is $v(x,\omega)=v_1(x) + v_2(x,\omega),$ such that $v_1\in L^\infty$ and $v_2$ is Gaussian with finite mean and variance.
For a measurable and integrable function $f$ on $\Omega,$ we define the expectation value of $f$ as
\begin{equation*}
{\mathbb E}(f):= \int f(\omega) \mu(d\omega).
\end{equation*}

We consider the $N$-body random Schr\"odinger operator 
\begin{equation}
H^N \equiv H_\omega^N := -\sum_{i=1}^N \Delta_i + \frac{1}{N}\sum_{1\le i<j \le N} v(x_i-x_j, \omega),
\end{equation}
where 
$\Delta = \sum_{j=1}^3 \frac{\partial^2}{\partial x_j^2}$ is the 3-dimensional Laplacian and $\omega\in\Omega.$ Here, we work in units where Planck's constant $\hbar=1$ and the mass of each particle is $m=\frac{1}{2}.$ We note that the analysis below is uniform in $\hbar.$  The Hamiltonian $H^N$ acts on the Hilbert space ${\mathcal H}^{N}:= L^2_S({\mathbb R}^{3N}),$ the symmetrization of $L^2({\mathbb R}^{3N})$, which is the space of pure states for a system of $N$ nonrelativistic bosons. 

The quantum dynamics  of the $N$-body system is  described by the Schr\"odinger equation
\begin{equation}
\label{eq:schrodinger}
i \partial_t\,\Psi^N(t)=H^N\Psi^N(t),
\end{equation}
with an initial condition 
$\Psi^N(t=0)=\Psi^{N,0}\in L_{S}^2({\mathbb R}^{3N})$. 

Together with the dynamics defined above, the $N$-body system is described by a kinematical algebra of ``observables''. For $p\leq N$, a $p$-particle observable is described by an operator $a^{(p)}\in {\mathcal B}({\mathcal H}^{(p)}),$ where ${\mathcal B}({\mathcal H}^{(p)})$ is the algebra of bounded operators on ${\mathcal H}^{(p)}=L^2_S({\mathbb R}^{3p}).$ By the nuclear theorem, one can associate with $\ap$ a tempered distribution kernel in ${\mathcal S}^\prime({\mathbb R}^{3p}\times {\mathbb R}^{3p})$, $\alpha^{(p)}(x_1,\ldots,x_p;y_1,\ldots,y_p):=\alpha^{(p)}(X_p;Y_p),$ such that 
\begin{equation}
\label{eq:kernel}
(a^{(p)}\varphi^{(p)})(X_p)=\int_{{\mathbb R}^{3p}}\alpha^{(p)} (X_p;Y_p)\varphi^{(p)}(Y_p)\,dY_p
\end{equation}
where $\varphi^{(p)}(Y_p)\in L^2_S({\mathbb R}^{3p}).$ 
We associate to $\ap$ an operator  $A^N(a^{(p)})$ acting  on ${\mathcal H}^{(N)}$ that is given by
\begin{equation}
(A^{N}(a^{(p)})\Psi)(x_1,\cdots,x_N)=\frac{N!}{N^p(N-p)!}
(P_{S}a^{(p)} \otimes I^{(N-p)} P_{S} \Psi)(x_1,\ldots,x_N), \label{eq:A^N}
\end{equation}
where $\Psi(x_1,\ldots,x_N)\in L^2_S({\mathbb R}^{3N})$ and $P_{S}$ is the projection onto the symmetric subspace  $L^{2}_S({\mathbb R}^{3N})$ of $L^{2}({\mathbb R}^{3N}).$ It follows from (\ref{eq:kernel}) and (\ref{eq:A^N}) that the map
$$A^N: {\mathcal B}({\mathcal H}^{(p)})\rightarrow  {\mathcal B}({\mathcal H}^{(N)}),  \ \ 1\le p\le N,$$ 
is linear, such that 
\begin{align}
& \| A^N(a^{(p)}) \|_{ {\mathcal B}({\mathcal H}^{(N)})} \le \|a^{(p)}\|_{{\mathcal B}( {\mathcal H}^{(p)})} ,\label{eq:ObservableNormBound}\\
& A^N(a^{(p)})^* = A^N(a^{(p) *}).\nonumber
\end{align}

In the Heisenberg picture, the evolution of $A^{(N)}\in {\mathcal B}({\mathcal H}^{(N)})$ is given by
\begin{equation}
\label{eq:evolution}
\alpha^N_{t}(A^{(N)}):= e^{iH^N t} A^{(N)} e^{-iH^Nt}, \ \ t\in \bbR.
\end{equation}
Since $v$ is almost surely bounded, $H^N$ is almost surely self-adjoint on the symmetrized Sobolev space $H^2_S({\mathbb R}^{3N}),$  and hence the propagator $e^{-iH^N t}, \ \ t\in {\mathbb R},$ is almost surely unitary. Moreover, it follows from the fact that the pointwise limit of measurable functions is itself measurable, \cite{Billing}, and the Trotter product formula, \cite{RS}, that 
$$\langle \otimes_{j=1}^N \psi_j (x_j), \alpha^N_t(A^{(N)}) \otimes_{j=1}^N \psi_j(x_j)\rangle, \ \ A^{(N)}\in {\mathcal B}({\mathcal H}^{(N)}), \ \ \psi_j \in L^2({\mathbb R}^{3})$$ is $\omega$-measurable. 

We now introduce the {\it classical} evolution. The Hartree equation is given by 
\begin{equation}
\label{eq:Hartree}
i\partial_t \psi_t = -\Delta \psi_t + (v\star |\psi_t|^2) \psi_t,
\end{equation}
with the initial condition $\psi_{t=0}=\phi \in L^2 (\bbR^3).$ It follows from Duhamel's formula for $\psi_t$ and the fact that $v\in L^\infty$ almost surely, that global solutions of (\ref{eq:Hartree}) in $L^2$ exist almost surely, such that 
$\|\psi_t \|_{L^2} = \| \phi\|_{L^2}$ with probability 1, for all $t\in \bbR,$ (see for example \cite{Ca} for the case when $v\in L^\infty$). It also follows from Duhamel's formula that the random variable 
\begin{equation*}
\langle \otimes_{i=1}^p \psi_t, A^{(p)} \otimes_{i=1}^p \psi_t\rangle , \ \ A^{(p)}\in\cB (\cH^{(p)}) 
\end{equation*}
is $\omega$-measurable.

\subsection{Statement of the main result}

We are in a position to state the main result. 

\begin{theorem}\label{th:Main}
 Given $a^{(p)},$ $A^N(a^{(p)})$ and $\alpha^N_t$ as above, suppose that the initial state of the $N$-body system is a normalized coherent (product) state $\Psi^{N,0}(x_1,\cdots,x_N)=\otimes_{i=1}^N \phi(x_i), \ \ \phi\in L^2({\mathbb R}^3).$ Then, for fixed $t\ge 0,$
\begin{equation*}
\lim_{N\rightarrow \infty}{\mathbb E}(\langle \Psi^{N,0}, \alpha^{N}_t (A^N(a^{(p)})) \Psi^{N,0}\rangle) = {\mathbb E} (\langle \otimes_{i=1}^p \psi_t, a^{(p)} \otimes_{i=1}^p \psi_t\rangle),
\end{equation*}
where $\psi_t$ satisfies the Hartree equation (\ref{eq:Hartree}) with initial condition $\psi_{t=0}=\phi.$
\end{theorem}

We note that the analysis below can be easily extended to study the mean-field dynamics of bosons in a random external potential that is almost surely smooth, polynomially bounded and positive, and to investigate the semi-classical limit of the dynamics under additional assumptions on the decay of the interaction, as in \cite{FGS}. Furthermore, the analysis below can be applied ``in toto'' to extend the results of \cite{FKP} and \cite{FKL} to the case of random interactions.

%%%%%%%%%%%%%%%%%%%%%%%%%%%%%%%%%%%%%%%%%%%%%%%%%%%%%

%\vspace{0.5cm}
%\noindent {\bf Acknowledgements}. The partial financial support of NSERC grant NA 7901 is gratefully acknowledged. 

%%%%%%%%%%%%%%%%%%%%%%%%%%%%%%%%%%%%%%%%%%%%%%%%%%%%%

\section{Proof of Theorem \ref{th:Main}}

The proof of Theorem \ref{th:Main} follows effectively from an application of the dominated convergence theorem, see \cite{Billing}, and Theorem 1.1 in \cite{FGS}. In what follows, we drop the explicit dependence on the time $t$ in the notation, since we fix it.

\begin{proof}

We introduce the random variables 
\begin{equation*}
\XN := \langle \Psi^{N,0}, \alpha^{N}_t (A^N(a^{(p)})) \Psi^{N,0}\rangle
\end{equation*}
and 
\begin{equation*}
\X := \langle \otimes_{i=1}^p \psi_t, a^{(p)} \otimes_{i=1}^p \psi_t\rangle .
\end{equation*}
The claim of the theorem is equivalent to the statement 
\begin{equation}
\label{eq:claim}
\lim_{N\rightarrow\infty} {\mathbb E}(\XN) = {\mathbb E}(\X).
\end{equation}

We divide the proof of (\ref{eq:claim}) into several steps.

\noindent {\it Step 1. Uniform integrability.}
We want to show that 
\begin{equation}
\label{eq:RVNBd3}
\lim_{\beta \rightarrow\infty} {\mathbb E} (|\XN| {\mathbf 1}_{|\XN|\ge \beta}) = 0, 
\end{equation}
uniformly in $N\in {\mathbb N}$. 

We have from (\ref{eq:ObservableNormBound}) and the fact that the quantum time-evolution is almost surely unitary, that
\begin{equation}
\label{eq:RVNBd}
|\XN | \le \| \ap \|_{\cB (\cH^{(p)} ) } < 2 \| \ap \|_{\cB (\cH^{(p ) }) } <\infty , \ \ \mathrm{almost \ \ surely},
\end{equation}
uniformly in $N\in {\mathbb N}.$
For $\beta > 0,$ it follows from (\ref{eq:RVNBd}) that 
\begin{equation}
\label{eq:RVNBd2}
|\XN | {\mathbf 1}_{|\XN|\ge \beta} \le |\XN| <  2 \| \ap \|_{\cB (\cH^{(p)})} <\infty , \ \ \mathrm{almost \ \ surely},
\end{equation}
uniformly in $N\in {\mathbb N}.$ The dominated convergence theorem together with (\ref{eq:RVNBd2}) give (\ref{eq:RVNBd3}).

\noindent {\it Step 2. Mean-field limit with probability 1.}
It follows from the fact that the particle interaction $v\in L^\infty$ almost surely and Theorem 1.1 in \cite{FGS} that, for fixed $t>0,$ 
\begin{equation}
\label{eq:ASMeanField}
\XN \stackrel{N\rightarrow\infty}{\rightarrow} \X \ \ \mathrm{almost \ \ surely}.
\end{equation} 

\noindent {\it Step 3.}
It follows from Fatou's lemma, \cite{Billing}, and (\ref{eq:RVNBd}), that 
\begin{equation}
\label{eq:XBd}
{\mathbb E} (|\X |) \le \liminf_{N} {\mathbb E} (|\XN |) \le \limsup_{N} {\mathbb E} (|\XN |) < 2 \| \ap \|_{\cB (\cH^{(p)})} <\infty ,
\end{equation}
uniformly in $N\in {\mathbb N}.$ We also have that  
\begin{equation*}
|\X| {\mathbf 1}_{|\X|\ge \beta} \le |\X|,
\end{equation*}
which together with (\ref{eq:XBd}) and the dominated convergence theorem, imply that 
\begin{equation}
\label{eq:XBd2}
\lim_{\beta \rightarrow\infty} {\mathbb E} (|\X| {\mathbf 1}_{|\X|\ge \beta}) = 0 .
\end{equation}

\noindent {\it Step 4. Convergence as $N\rightarrow\infty.$}
We introduce the random variable 
\begin{equation*}
\YN := |\X - \XN|.
\end{equation*}
Note that it suffices to show that ${\mathbb E}(\YN)\rightarrow 0$ as $N\rightarrow\infty$, from which (\ref{eq:claim}) follows by the triangular inequality.

It follows from (\ref{eq:ASMeanField}), Step 2, that 
\begin{equation}
\label{eq:ASConv}
\YN \stackrel{N\rightarrow\infty}{\rightarrow} 0 \ \ \mathrm{almost \ \ surely.}
\end{equation}

We decompose $\YN$ into two parts,
\begin{equation*}
\YN = \YNl + \YNg,
\end{equation*}
where $\YNl := \YN {\mathbf 1}_{|\YN| <\beta}$ and $\YNg := \YN {\mathbf 1}_{|\YN| \ge \beta},$ for $\beta >0.$ 

Since $\YNl <\beta,$ (\ref{eq:ASConv}) together with the dominated convergence theorem imply that 
\begin{equation}
\label{eq:LimL}
\lim_{N\rightarrow\infty } {\mathbb E}(\YNl) = 0.
\end{equation}
Furthermore, since 
$$\YNg \le 2|\X|{\mathbf 1}_{|\X|\ge\beta/2}+2|\XN|{\mathbf 1}_{|\XN|\ge\beta/2},$$
it follows from (\ref{eq:RVNBd3}) and (\ref{eq:XBd2}) that 
\begin{equation}
\label{eq:LimG}
\lim_{\beta \rightarrow\infty } {\mathbb E}(\YNg) = 0,
\end{equation}
uniformly in $N\in {\mathbb N}.$

Given $\epsilon >0,$ (\ref{eq:LimG}) implies that there exists a finite $\beta_0 >0$ such that 
\begin{equation*}
\sup_{N}{\mathbb E}(Y_N^{(p),\ge \beta_0}) <\epsilon/2 .
\end{equation*}
Moreover, (\ref{eq:LimL}) implies that there exists a positive integer $N_0$ such that, for all $N\ge N_0,$
\begin{equation*}
{\mathbb E}(Y_N^{(p),<\beta_0}) <\epsilon/2.
\end{equation*}
It follows that 
\begin{equation*}
{\mathbb E}(\YN) = {\mathbb E}(Y_N^{(p),<\beta_0}) + {\mathbb E}(Y_N^{(p),\ge \beta_0}) < \epsilon
\end{equation*}
for $N\ge N_0.$ Therefore, ${\mathbb E}(\YN)\stackrel{N\rightarrow\infty}{\rightarrow} 0.$

By the triangular inequality, 
\begin{equation*}
|{\mathbb E}(\X) - {\mathbb E} (\XN)| \le {\mathbb E}(\YN)\stackrel{N\rightarrow\infty}{\rightarrow} 0 ,
\end{equation*}
which gives the claim of the theorem. \end{proof}

%%%%%%%%%%%%%%%%%%%%%%%%%%%%%%%%%%%%%%%%%%%%%%%%%%%%%%
%BIBLIOGRAPHY %%%%%%%%%%%%%%%%%%%%%%%%%%%%%%%%%%%%%%%%
%%%%%%%%%%%%%%%%%%%%%%%%%%%%%%%%%%%%%%%%%%%%%%%%%%%%%%

\end{document}